\documentclass[12pt]{article}

\usepackage{epsfig,epic,eepic}
\usepackage{graphicx}
\usepackage[all]{xy}

\usepackage{booktabs}
\usepackage{color}
\usepackage{algorithm}
\usepackage[noend]{algorithmic}
\usepackage{stmaryrd}
\usepackage{amsfonts}
\usepackage{amsmath, amsthm, amssymb}

\newcommand{\B}{\mathbb{B}}

\newtheorem{theorem}{Theorem}[section]
\newtheorem{corollary}[theorem]{Corollary}
\newtheorem{lemma}[theorem]{Lemma}

\newtheorem{fact}[theorem]{Fact}

\theoremstyle{remark}

\theoremstyle{example}
\newtheorem{example}[theorem]{Example}

\theoremstyle{definition}
\newtheorem{definition}[theorem]{Definition}

\usepackage{listings}
\lstdefinelanguage{PseudoC}[ISO]{C++} { 
morekeywords={foreach, and, not, or, is, FIFO_Queue, HashTable, FILE, Cache}, 
morekeywords=[2]{HashInsert, Enqueue, Dequeue, next}, 
mathescape=true,
alsoletter={'},
deletestring=[b]'
}

\lstdefinelanguage{PseudoCWithNumbers}[ISO]{C++} { 
morekeywords={foreach, and, not, or, is, FIFO_Queue, HashTable, FILE, Cache}, 
morekeywords=[2]{HashInsert, Enqueue, Dequeue, next}, 
mathescape=true,
alsoletter={'},
deletestring=[b]'
}

\lstdefinelanguage{Murphi}[]{Pascal} { 
morekeywords={ruleset, rule, invariant, startstate, return, endif, endfor, endswitch, forall, endforall, exists, endexists}, 
mathescape=true, 
morestring=[b]",
morestring=[b]', 
morecomment=[s]{/*}{*/} ,
morecomment=[l]{--}
}

\lstdefinelanguage{PRISM}[ISO]{C++} { 
morekeywords={probabilistic, stochastic, const, rate, module, endmodule, init, P, U},
mathescape=true, 
alsoletter={'},
deletestring=[b]'
}

\lstdefinelanguage{Yacc}[ISO]{C++} { 
morekeywords={token, left}, 
mathescape=false,
alsoletter={'},
deletestring=[b]'
}

\lstdefinestyle{PseudoC}{
language=PseudoC,
basicstyle=\ttfamily,
tabsize=1,
showlines=false,
emptylines=*1,
breaklines=true,
breakindent=5pt,
keywordstyle=\rmfamily\bfseries,
keywordstyle=[2]\rmfamily,
commentstyle=\itshape, 
columns=fixed,
showspaces=false, 
showstringspaces=false, 
showtabs=false, 
escapechar=\%
}

\lstdefinestyle{PseudoCWithNumbers}{
language=PseudoC,
basicstyle=\ttfamily,
tabsize=1,
showlines=false,
emptylines=*1,
breaklines=true,
breakindent=5pt,
keywordstyle=\rmfamily\bfseries,
keywordstyle=[2]\rmfamily,
commentstyle=\itshape, 
columns=fixed,
showspaces=false, 
showstringspaces=false, 
showtabs=false, 
escapechar=\%,
numbersep=5pt,framexleftmargin=15pt,numbers=left,
}

\lstdefinestyle{Murphi}{
language=Murphi,
basicstyle=\ttfamily,
tabsize=1,
showlines=false,
emptylines=*1,
breaklines=true,
breakindent=5pt,
keywordstyle=\rmfamily\bfseries,
commentstyle=\itshape, 
columns=fixed,
showspaces=false, 
showstringspaces=false, 
showtabs=false,
escapechar=\%}

\lstdefinestyle{PRISM}{
language=PRISM,
basicstyle=\ttfamily,
tabsize=1,
showlines=false,
emptylines=*1,
breaklines=true,
breakindent=5pt,
keywordstyle=\rmfamily\bfseries,
commentstyle=\itshape, 
columns=fixed,
showspaces=false, 
showstringspaces=false, 
showtabs=false,
escapechar=\%}

\lstdefinestyle{Yacc}{
language=PseudoC,
basicstyle=\ttfamily,
tabsize=1,
showlines=false,
emptylines=*1,
breaklines=true,
breakindent=5pt,
keywordstyle=\rmfamily\bfseries,
keywordstyle=[2]\rmfamily,
commentstyle=\itshape, 
columns=fixed,
showspaces=false, 
showstringspaces=false, 
showtabs=false, 
}

\newcommand{\myboldsub}[1]{\mbox{\scriptsize\boldmath$#1$}}
\newcommand{\mybold}[1]{\mbox{\boldmath$#1$}}
\newcommand{\fun}[1]{{\textsl{#1}}}
\newcommand{\dotcup}{\ensuremath{\mathaccent\cdot\cup}}
\definecolor{light-gray}{gray}{0.90}
\newcommand{\qks}{\mbox{QKS}}

\title{From Boolean Functional Equations\\to Control Software}

\author{Federico Mari, Igor Melatti, Ivano Salvo, Enrico Tronci\\
\small \itshape Department of Computer Science\\
\small \itshape Sapienza University of Rome\\
\small \itshape via Salaria 113, 00198 Rome\\
\small email: \{mari,melatti,salvo,tronci\}@di.uniroma1.it}

\begin{document}

\maketitle

\begin{abstract}

Many software as well digital hardware automatic synthesis methods define the
set of implementations meeting the given system specifications with a boolean
relation $K$. In such a context a fundamental step in the software (hardware)
synthesis process is finding effective solutions to the functional equation
defined by $K$.  This entails finding a (set of) boolean function(s) $F$ 
(typically represented using OBDDs, \emph{Ordered Binary Decision Diagrams})
such that: 1) for all $x$ for which $K$ is satisfiable, $K(x, F(x)) = 1$ holds;
2) the implementation of $F$ is efficient  with respect to given implementation
parameters such as code size or execution time.
While this problem has been widely studied in digital hardware synthesis, little
has been done in a software synthesis context. 
Unfortunately the approaches developed for hardware synthesis cannot be directly
used in a software context. This motivates investigation of effective methods to
solve the above problem when $F$ has to be implemented with software.

%
In this paper we present an algorithm that, from an OBDD representation for $K$,
generates a C code implementation for $F$ that has the same size as the OBDD for
$F$ and a WCET  (\emph{Worst Case Execution Time}) at most $O(nr)$, being $n =
|x|$ the number of input arguments for functions in $F$ and $r$ the number of
functions in $F$. 

\end{abstract}

\section{Introduction}\label{intro.tex}


Many software as well digital hardware automatic synthesis methods define the
set of implementations meeting the given system specifications with a boolean
relation $K$. Such relation typically takes as input (the $n$-bits encoding of)
a {\em state} $x$ of the system and (the $r$-bits encoding of) a proposed {\em
action} to be performed $u$, and returns {\em true} (i.e. $1$) iff the system
specifications are met when performing action $u$ in state $x$. In such a
context a fundamental step in the software (hardware) synthesis process is
finding effective solutions to the functional equation defined by $K$, i.e.
$K(x, u) = 1$. This entails finding a tuple of boolean functions $F = \langle
f_1, \ldots, f_r \rangle$  (typically represented using OBDDs, \emph{Ordered
Binary Decision Diagrams}~\cite{Bry86}) s.t. 1) for all $x$ for which $K$ is
satisfiable (i.e., it enables at least one action), $K(x, F(x)) = 1$ holds, and
2) the implementation of $F$ is efficient  with respect to given implementation
parameters such as code size or execution time.

While this problem has been widely studied in digital hardware
synthesis~\cite{BCK09}, little has been done in a software synthesis context.
This is not surprising since software synthesis from formal specifications is
still in its infancy. Unfortunately the approaches developed for hardware
synthesis cannot be directly used in a software context.  In fact, synthesis
methods targeting a hardware implementation typically aim at minimizing the
number of digital gates and of hierarchy levels. Since in the same hierarchy
level gates output computation  is {\em parallel}, the hardware implementation
WCET (\emph{Worst Case Execution Time}) is given by the number of levels. On the
other hand, a software implementation will have to {\em sequentially} compute
the gates outputs. This implies that the software implementation WCET is the
number of gates used, while a synthesis method targeting a software
implementation may obtain a better WCET. This motivates investigation of
effective methods to solve the above problem when $F$ has to be implemented with
software.


\subsection{Our Contribution}\label{contribution.subsec}


In this paper we present an algorithm that, from an OBDD representation for $K$,
effectively generates a C code implementation for $K$. This is done in two
steps:

\begin{enumerate}

	\item \label{cobdd_to_f.lbl} from an OBDD representation for $K$ we
effectively compute an OBDD representation for $F$, following the lines
of~\cite{Tro98}; 

	\item \label{f_to_c.lbl} we generate a C code implementation for $F$
that has the same size as the OBDD for $F$ and a $O(nr)$ WCET,
being $n = |x|$ the size of states encoding and $r = |u|$ the size of actions
encoding. Indeed, we prove a more strict upper bound for the WCET by also
considering the heights of the OBDDs representing $F$.

\end{enumerate}

We formally prove both steps~\ref{cobdd_to_f.lbl} and~\ref{f_to_c.lbl} to be 
correct. This allows us to synthesize correct-by-construction {\em control
software}, provided that $K$ is provably correct w.r.t. initial formal
specifications. This is the case of~\cite{qks-cav2010}, where an algorithm to
synthesize $K$ starting from the formal specification of a Discrete-Time Linear
Hybrid System ({\em DTLHS} in the following) is presented. Thus this methodology
allows a  correct-by-construction control software to be synthesized, starting
from formal specifications for DTLHSs.

Note that the problem of solving the functional equation $K(x, F(x)) = 1$ w.r.t.
$F$ is trivially decidable, since there are finitely many $F$. However, trying
to explicitly enumerate all $F$ requires time $\Omega(2^{r2^n})$ (being $n$ the
number of bits encoding state $x$ and $r$ the number of bits encoding state
$u$). By using OBDD-based computations, our algorithm complexity is $O(r2^n)$ in
the worst case. However, in many interesting cases OBDD sizes and computations
are much lower than the theoretical worst case (e.g. in Model Checking
applications, see~\cite{CGP99}).


Furthermore, once the OBDD representation for $F$ has been computed, a trivial
implementation of $F$ could use a look-up table in RAM. While this solution
would yield a better  WCET, it would imply a $\Omega(r2^n)$ RAM usage.
Unfortunately, implementations for $F$ in real-world cases are typically
implemented on microcontrollers (this is the case e.g. for {\em embedded
systems}). Since microcontrollers usually have a small RAM, the look-up table
based solution is not feasible in many interesting cases. The approach we
present here only requires $O(n + r)$ bytes of RAM for the data. As for the
program size, it is linear in the size (i.e., number of nodes) of the OBDDs
representing $F$, thus again we rely on the compression OBDDs achieve in many
interesting cases.

Moreover, $F : \B^n \to \B^r$ is composed by $r$ boolean functions, thus it is
represented by $r$ OBDDs. Such OBDDs typically share nodes among them. If a 
trivial implementation of $F$ in C code is used, i.e. each OBDD is translated as
a stand-alone C function, OBDDs nodes sharing will not be exploited. In our
approach, we also exploit nodes sharing, thus the control software we generate
fully takes advantage of OBDDs compression.

Finally, we present experimental results showing effectiveness of the proposed
algorithm. As an example, in less than 1 second and within 70 MB of RAM we are
able to synthesize the control software for a function $K$ of $24$ boolean
variables, divided in $n=20$ state variables and $r=4$ action variables,
represented by a OBDD with about $4 \times 10^4$ nodes. Such $K$ represents the
set of correct implementations for a real-world system, namely a multi-input
buck DC/DC converter~\cite{buck-tekrep-2011}, obtained as described
in~\cite{qks-cav2010}. The control software we synthesize in such a case has
about $1.2 \times 10^4$ lines of code, whilest a control software not taking
into account OBDDs nodes sharing would have had about $1.5 \times 10^4$ lines of
code. Thus, we obtain a $24\%$ gain towards a trivial implementation.


\subsection{Related Work}\label{related_work.subsec}


Synthesis of boolean functions $F$ satisfying a given boolean relation $K$ in a
way s.t. $K(x, F(x)) = 1$ is also addressed in~\cite{BCK09}.
However,~\cite{BCK09} targets a hardware setting, whereas we are interested in a
software implementation for $F$. Due to structural differences between hardware
and software based implementations (see the discussion above), the method
in~\cite{BCK09} is not directly applicable here.

In~\cite{qks-cav2010} an algorithm is presented which, starting from formal
specifications of a DTLHS, synthesizes a correct-by-construction boolean
relation $K$, and then a correct-by-construction control software implementation
for $K$. However, in~\cite{qks-cav2010} the implementation of $K$ is neither
described in detail, nor it is proved to be correct. Furthermore, the
implementation synthesis described in~\cite{qks-cav2010} has not the same size
of the OBDD for $F$, i.e. it does not exploit OBDD node sharing.

In~\cite{Tro98} an algorithm is presented which computes boolean functions $F$
satisfying a given boolean relation $K$ in a way s.t. $K(x, F(x)) = 1$. This
approach is very similar to ours. However~\cite{Tro98} does not generate the C
code control software and it does not exploit OBDD node sharing. Furthermore,
the algorithm is not proved to be correct.

Therefore, to the best of our knowledge this is the first time that an algorithm
synthesizing correct-by-construction control software starting from a boolean
relation (with the characteristics given in Sect.~\ref{contribution.subsec}) is
presented and proved to be correct.

%
%

\section{Basic Definitions}\label{basic.tex}

In the following, we denote with $\B = \{0, 1\}$ the boolean domain, where $0$
stands for {\em false} and $1$ for {\em true}. We will denote boolean functions
$f : \B^n \to \B$ with boolean expressions on boolean variables involving $+$
(logical OR), $\cdot$ (logical AND, usually omitted thus $xy = x \cdot y$),
$\bar{\ }$ (logical complementation) and $\oplus$ (logical XOR). We also denote
with $f|_{x_i = g}(x_1, \ldots, x_n)$ the boolean function $f(x_1, \ldots, x_{i
- 1}, g(x_1, \ldots, x_n), x_{i + 1}, \ldots, x_n)$ and with $\exists x_i\;
f(x_1, \ldots, x_n)$ the boolean function $f|_{x_i = 0}(x_1, \ldots, x_n) +
f|_{x_i = 1}(x_1, \ldots, x_n)$. We will also denote vectors of boolean
variables in boldface, e.g. $\mybold{x} = \langle x_1, \ldots, x_n\rangle$.

Finally, we denote with $[n]$ the set $\{1, \ldots, n\}$.


\subsection{Feedback Control Problem for Labeled Transition Systems}
\label{lts.tex}


In this paper we focus on solving and implementing a functional equation
$K(\mybold{x}, \mybold{u}) = 1$. In this section we show a typical case in which
such an equation needs to be solved and implemented. 

A \emph{Labeled Transition System} (LTS) is a tuple ${\cal S} = (S, A, T)$
where  $S$ is a finite set of states,  $A$ is a finite
set of \emph{actions}, and  $T$ : $S$ $\times$ $A$ $\times$ $S$ $\to$
$\B$ is the \emph{transition relation} of ${\cal S}$.
An LTS is {\em deterministic} if $T(s, a, s') \land T(s, a, s'') \Rightarrow s'
= s''$, and {\em nondeterministic} otherwise.
A {\em run} or \emph{path} for an LTS ${\cal S}$  is a sequence  $\pi$ = $s_0,
a_0, s_1, a_1, s_2, a_2, \ldots$  of states $s_t$ and actions $a_t$  such that
$\forall t \geq 0$ $T(s_t, a_t, s_{t+1})$. The length $|\pi|$ of a finite run
$\pi$ is the number of actions in $\pi$. We denote with $\pi^{(S)}(t)$ the $t$-th
state element of $\pi$.

A \emph{controller} for an LTS ${\cal S}$ is a function $K : S \times A \to \B$
such that $\forall s \in S$, $\forall a \in A$, if $K(s, a) = 1$ then $\exists
s' \in S \; T(s, a, s') = 1$. We denote with $\mbox{\rm Dom}(K)$ the set of
states for which a control action is defined. Formally, $\mbox{\rm Dom}(K)$ $=$
$\{s \in S \; | \; \exists a \; K(s, a)\}$. ${\cal S}^{(K)}$ denotes the
\emph{closed loop system}, that is the LTS $(S, A, T^{(K)})$, where $T^{(K)}(s,
a, s') = T(s, a, s') \wedge K(s, a)$. 

In the following, by assuming proper boolean encoding functions for states and
actions (as it is usually done in Model Checking applications,
see~\cite{CGP99}), we may see a controller as a boolean function $K: \B^n \times
\B^r \to \B$, with $n = \lceil\log_2 |S|\rceil$ and $r = \lceil\log_2
|A|\rceil$. 


We call a path $\pi$ {\em fullpath} \cite{emerson-toplas-04}  if either it is
infinite or its last state  $\pi^{(S)}(|\pi|)$  has no successors (i.e.
$\mbox{\rm Adm}({\cal S}, \pi^{(S)}(|\pi|)) = \varnothing$). We denote with
${\rm Path}(s)$ the set of fullpaths starting in state $s$, i.e. the set of
fullpaths $\pi$ such that $\pi^{(S)}(0)=s$.

Given a path $\pi$ in ${\cal S}$, we define the measure $J({\cal S},G,\pi)$ on
paths as the distance of $\pi^{(S)}(0)$ to the goal on $\pi$. That is,  if there
exists $n > 0$ s.t. $\pi^{(S)}(n)\in G$, then $J({\cal S},\pi,G)$ $=$ $\min\{n$
$|$ $n > 0 \land \pi^{(S)}(n)\in G\}$. Otherwise, $J({\cal S},\pi,G) = +\infty$.
We require $n > 0$ since our systems are nonterminating and each controllable
state (including a goal state) must have a path of positive length to a goal
state. The {\em worst case distance} (pessimistic view) of a state $s$ from the
goal region  $G$ is  $J_{\rm strong}({\cal S},G,s)=\sup \{ J({\cal S},G,s,
\pi)~|~ \pi \in{\rm Path}(s)\}$. 

\begin{definition}
\label{def:sol}
Let ${\cal P}$ = $({\cal S}$, $I$, $G)$ 
be a control problem and $K$ be  
a controller  for ${\cal S}$ such that 
$I$ $\subseteq$ $\mbox{\rm Dom}(K)$.

$K$ is a {\em strong} solution to ${\cal P}$ 
if for all $s \in \mbox{\rm Dom}(K)$, 
$J_{\rm strong}({\cal S}^{(K)}, G, s)$ 
is finite.

%
An \emph{optimal} strong solution to ${\cal P}$
is a strong solution $K^{*}$ to ${\cal P}$ such that 
for all strong solutions
$K$ to ${\cal P}$, for all $s \in S$ we have: 
$J_{\rm strong}({\cal S}^{(K^{*})}, G, s) 
\leq J_{\rm strong}({\cal S}^{(K)}, G, s)$.

\end{definition}

Intuitively, a strong solution takes a \emph{pessimistic} view and
requires that for each initial state, \emph{all} runs in the closed
loop system reach the goal (no matter nondeterminism outcomes).
Unless otherwise stated, we call just solution a
strong solution.

\begin{definition}
\label{def:mgo}

%
The  \emph{most general optimal (mgo) strong solution} (simply {\em mgo} in the
following) to ${\cal P}$ is an optimal strong solution  $\bar{K}$ to ${\cal P}$
such that   for all other optimal strong solutions $K$ to ${\cal P}$, for all $s
\in S$, for all $a \in A$ we have that $K(s, a)$ $\Rightarrow$ $\bar{K}(s, a)$. 

\end{definition}

Efficient algorithms to compute mgos starting from suitable (nondeterministic)
LTSs have been proposed in the literature (e.g. see~\cite{strong-planning-98}).
Once an mgo $K$ has been computed, solving and implementing the functional
equation $K(\mybold{x}, \mybold{u}) = 1$ allows a  correct-by-construction
control software to be synthesized.

\subsection{OBDD Representation for Boolean Functions}\label{obdd.subsec}


A {\em Binary Decision Diagram} (BDD) $R$ is a rooted directed acyclic graph
(DAG) with the following properties. Each $R$ node $v$ is labeled either with a
boolean variable ${\rm var}(v)$ (internal node) or with a boolean constant 
${\rm val}(v) \in \B$ (terminal node). Each $R$ internal node $v$ has exactly
two children, labeled with ${\rm high}(v)$ and ${\rm low}(v)$. Let $x_1, \ldots,
x_n$  be the boolean variables labeling $R$ internal nodes. Each terminal node
$v$  represents the (constant) boolean function $f_v(x_1, \ldots, x_n) = {\rm
val}(v)$. Each internal node $v$ represents the boolean function $f_v(x_1,
\ldots, x_n) = x_if_{{\rm high}(v)}(x_1, \ldots, x_n) + \bar{x}_if_{{\rm
low}(v)}(x_1, \ldots, x_n)$, being $x_i = {\rm var}(v)$. 

An {\em Ordered BDD} (OBDD) is a BDD where, on each path from the root to a
terminal node, the variables labeling each internal node must follow the same
ordering. Two OBDDs are {\em isomorphic} iff there exists a mapping from nodes
to nodes preserving attributes ${\rm var}$, ${\rm val}$, ${\rm high}$ and ${\rm low}$. 

An OBDD is called {\em reduced} iff it contains no vertex $v$ with ${\rm low}(v)
= {\rm high}(v)$, nor does it contain distinct vertices $v$ and $v'$ such that
the subgraphs rooted by $v$ and $v'$ are isomorphic. This entails that
isomorphic subgraphs are {\em shared}, i.e. only one copy of them is effectively
stored (see~\cite{Bry86}).

We will only deal with reduced OBDDs, thus we will call them simply OBDDs. It
can be shown~\cite{Bry86} that each boolean function can be represented by
exactly one OBDD (up to isomorphism), thus OBDD representation for boolean
functions is {\em canonical}.

\section{Solving a Boolean Functional Equation}\label{problem.tex}


Let $K(x_1, \ldots, x_n, u_1, \ldots, u_r)$ be an mgo for a given control
problem ${\cal P} = ({\cal S},$ $I,$ $G)$. We want to solve the {\em boolean
functional equation} $K(\mybold{x}, \mybold{u}) = 1$ w.r.t. variables
$\mybold{u}$, that is we want to obtain boolean functions $f_1, \ldots, f_r$
s.t. $K(\mybold{x}, f_1(\mybold{x}), \ldots, f_r(\mybold{x})) = K|_{u_1 =
f_1(\myboldsub{x}), \ldots, u_r = f_r(\myboldsub{x})}(\mybold{x}, \mybold{u}) =
1$. 

This problem may be solved in different ways, depending on the {\em target
implementation} (hardware or software) for functions $f_i$. In both cases, it is
crucial to be able to bound the WCET (\emph{Worst Case Execution Time}) of the
obtained controller. In fact, controllers must work in an endless closed loop
with the system ${\cal S}$ ({\em plant}) they control. This implies that, every
$T$ seconds ({\em sampling time}), the controller has to decide the actions to
be sent to the plant. Thus, in order for the entire system (plant + control
software) to properly work, the controller WCET upper bound must be at most
$T$. 

In~\cite{BCK09}, $f_1, \ldots, f_r$ are generated in order to optimize a {\em
hardware} implementation. In this paper, we focus on software implementations
for $f_i$ ({\em control software}). As it is discussed in Sect.~\ref{intro.tex},
simply translating an hardware implementation into a software implementation
would result in a too high WCET. Thus, a method directly targeting software is
needed. An easy solution would be to set up, for a given state $\mybold{x}$, a
SAT problem instance ${\cal C} = C_{K1}, \ldots, C_{Kt}, c_1, \ldots, c_n$,
where $C_{K1} \land \ldots \land C_{Kt}$ is equisatisfiable to $K$ and each
clause $c_i$ is either $x_i$ (if $x_i$ is $1$) or $\bar{x}_i$ (otherwise). Then
${\cal C}$ may be solved using a SAT solver, and the values assigned to
$\mybold{u}$ in the computed satisfying assignment may be returned as the action
to be taken. However, it would be hard to estimate a WCET for such an
implementation. The method we propose in this paper overcomes such obstructions
by achieving a WCET at most proportional to $rn$.


\section{OBDDs with Complemented Edges}\label{flipped-obdds.tex}

In this section we introduce OBDDs with complemented edges (COBDDs,
Def.~\ref{flipped-obdd.def}), which were first presented in~\cite{BRB90,MIY90}.
Intuitively, they are OBDDs where else edges (i.e. edges of type $(v, low(v))$)
may be complemented. Then edges (i.e. edges of type $(v, high(v))$)
complementation is not allowed to retain canonicity. Edge complementation
usually reduce resources usage, both in terms of CPU and memory.

\begin{definition}\label{flipped-obdd.def}

An {\em OBDD with complemented edges} (COBDD in the following) is a tuple $\rho
= ({\cal V}$, $V$, ${\bf 1}$, ${\rm var}$, ${\rm low}$, ${\rm high}$, ${\rm
flip})$ with the following properties:

\begin{enumerate}

	\item \label{variables.lbl} ${\cal V} = \{x_1, \ldots, x_n\}$ is a
	finite set of boolean variables s.t. for all $x_i \neq x_j \in {\cal
	V}$, either $x_i < x_j$ or $x_j < x_i$;

	\item \label{nodes.lbl} $V$ is a finite set of {\em nodes};

	\item \label{one.lbl} ${\bf 1} \in V$ is the {\em terminal} node of
	$\rho$, corresponding to the boolean constant $1$; any non-terminal node
	$v \in  V, v \neq {\bf 1}$ is called {\em internal};

	\item \label{labels.lbl} ${\rm var}, {\rm low}, {\rm high}, {\rm flip}$
are functions defined on internal nodes, namely: 

	\begin{itemize}
	
		\item ${\rm var}: V \setminus \{{\bf 1}\} \to {\cal V}$ assigns
		to each internal node a boolean variable in ${\cal V}$;

		\item ${\rm high} : V \setminus \{{\bf 1}\} \to V$ assigns to
		each internal node $v$ a {\em high child} (or {\em true child}),
		representing the case in which ${\rm var}(v) = 1$;

		\item ${\rm low} : V \setminus \{{\bf 1}\} \to V$ assigns to
		each internal node $v$ a {\em low child} (or {\em else child}),
		representing the case in which ${\rm var}(v) = 0$;

		\item ${\rm flip}: V \setminus \{{\bf 1}\} \to \B$ assigns to
		each internal node $v$ a boolean value; namely, if ${\rm
		flip}(v) = 1$ then the else child has to be complemented,
		otherwise it is regular (i.e. non-complemented);

	\end{itemize}

	\item \label{ordering.lbl} for each internal node $v$, ${\rm var}(v) <
	{\rm var}({\rm high}(v))$ and ${\rm var}(v) < {\rm var}({\rm low}(v))$.
	
\end{enumerate}

\end{definition}

\paragraph{COBDDs as (labeled) DAGs}\label{cobdd-dag.subsec}

A COBDD $\rho = ({\cal V}$, $V$, ${\bf 1}$, ${\rm var}$, ${\rm low}$, ${\rm
high}$, ${\rm flip})$ defines a labeled directed multigraph in a straightforward way.
This is detailed in Def.~\ref{flipped-obdd.graph.def}.

\begin{definition}\label{flipped-obdd.graph.def} 

Let $\rho = ({\cal V}$, $V$, ${\bf 1}$, ${\rm var}$, ${\rm low}$, ${\rm high}$,
${\rm flip})$ be a COBDD. The {\em graph associated to} $\rho$ is a labeled
directed multigraph $G^{(\rho)} = (V, E)$ where $V$ is the same set of nodes of $\rho$ and:

\begin{enumerate}

	\item \label{edges.lbl} $E = \{(v, w) \;|\; w = {\rm high}(v) \lor w
	= {\rm low}(v)\}$ ($E$ is a multiset since it may happen that ${\rm high}(v)
	= {\rm low}(v)$ for some $v \in V$);

	\item \label{labels-graph.lbl} the following labeling functions are
	defined on nodes and edges: 

	\begin{itemize}
	
		\item ${\rm ind}: V \setminus \{{\bf 1}\} \to {\cal V}$ assigns
		to each internal node $v$ a boolean variable in ${\cal V}$, and is
		defined by ${\rm ind}(v) = {\rm var}(v)$;

		\item ${\rm type} : E \to \{{\rm then}, {\rm else}, {\rm
		compl}\}$ assigns to each edge $e = (v, w)$ its type, and is
		defined by: ${\rm type}(e) = {\rm then}$ ({\em then edge}) iff
		${\rm high}(v) = w$, ${\rm type}(e) = {\rm else}$ ({\em regular
		else edge}) iff ${\rm low}(v) = w \land {\rm flip}(v) = 0$,
		${\rm type}(e) = {\rm compl}$ ({\em complemented else edge}) iff
		${\rm low}(v) = w \land {\rm flip}(v) = 1$.

	\end{itemize}
	
\end{enumerate}

\end{definition}

\begin{example}\label{cobdd.ex}

%
%
%
%
%
%
%
%
%

\sloppy

Let $\rho = (\{x_0$, $x_1$, $x_2\}$, $\{{\rm 0x15}$, ${\rm 0x14}$, ${\rm 0x13}$,
${\rm 0xe}$, ${\bf 1}\}$, ${\bf 1}$, ${\rm var}$, ${\rm low}$, ${\rm high}$,
${\rm flip})$ be a COBDD with: i) ${\rm var}({\rm 0x15}) = x_0$, ${\rm var}({\rm
0x14}) = {\rm var}({\rm 0x13}) = x_1$, ${\rm var}({\rm 0xe}) = x_2$ and $x_0 <
x_1 < x_2$; ii) ${\rm high}({\rm 0x15}) = {\rm 0x13}$, ${\rm low}({\rm 0x15}) =
{\rm 0x14}$, ${\rm high}({\rm 0x13}) = {\rm high}({\rm 0x14}) = {\rm 0xe}$,
${\rm high}({\rm 0xe}) =$ ${\rm low}({\rm 0xe}) =$ ${\rm low}({\rm 0x13}) =$
${\rm low}({\rm 0x14}) =$  ${\bf 1}$; iii) ${\rm flip}({\rm 0x14}) = 0$, ${\rm
flip}({\rm 0x15}) =$ ${\rm flip}({\rm 0x13}) =$ ${\rm flip}({\rm 0xe}) =$ $1$.

\fussy

%
%
%
%
%
%
%
%

Then $G^{(\rho)}$ is shown in Fig.~\ref{F1.example.eps}, where edges are
directed downwards. Moreover, in Fig.~\ref{F1.example.eps} then edges are solid
lines, regular else edges are dashed lines and complemented else edges are
dotted lines. 

\end{example}

\paragraph{Restriction of a COBDD}\label{restriction.subsec}


The graph associated to a given COBDD may be seen as a forest with multiple
rooted multigraphs. Def.~\ref{rooted-cobdd.def} allow us to select one root
vertex and thus one rooted multigraph.

\begin{definition}\label{rooted-cobdd.def}

Let $\rho = ({\cal V}$, $V$, ${\bf 1}$, ${\rm var}$, ${\rm low}$, ${\rm high}$,
${\rm flip})$ be a COBDD, and let $v \in V$. The {\em COBDD restricted to $v$}
is the COBDD $\rho_v = ({\cal V}$, $V_v$, ${\bf 1}$, ${\rm var}_v$, ${\rm
low}_v$, ${\rm high}_v$, ${\rm flip}_v)$  s.t.:

\begin{itemize}

	\item $V_v = \{w \in V \;|$ there exists a path from $v$ to $w$ in
	$G^{(\rho)}\}$ (note that $v \in V_v$);
	
	\item ${\rm var}_v$, ${\rm low}_v$, ${\rm high}_v$ and ${\rm flip}_v$
are the restrictions to $V_v$ of ${\rm var}$, ${\rm low}$, ${\rm high}$ and
${\rm flip}$.

\end{itemize}

\end{definition}

\paragraph{Reduced COBDDs}\label{reduction.subsec}

Two COBDDs are {\em isomorphic} iff there exists a mapping from nodes to nodes
preserving attributes ${\rm var}$, ${\rm flip}$, ${\rm high}$ and ${\rm low}$. A
COBDD is called {\em reduced} iff it contains no vertex $v$ with ${\rm low}(v) =
{\rm high}(v) \land {\rm flip}(v) = 0$, nor does it contains distinct vertices
$v$ and $v'$ such that $\rho_v$ and $\rho_{v'}$ are isomorphic. Note that,
differently from OBDDs, it is possible that ${\rm high}(v) = {\rm low}(v)$ for
some $v \in V$, provided that ${\rm flip}(v) = 1$ (e.g. see nodes ${\rm 0xf}$
and ${\rm 0xe}$ in Fig.~\ref{F2.example.eps}). In the following, we assume all
our COBDDs to be reduced. 

\paragraph{COBDDs Properties}\label{cobdd_properties.subsec}

For a given COBDD $\rho = ({\cal V}$, $V$, ${\bf 1}$, ${\rm var}$, ${\rm low}$,
${\rm high}$, ${\rm flip})$ the following properties follow from
definitions~\ref{flipped-obdd.def} and~\ref{flipped-obdd.graph.def}: i)
$G^{(\rho)}$ is a rooted directed acyclic (multi)graph (DAG); ii) each path in
$G^{(\rho)}$ starting from an internal node ends in ${\bf 1}$; iii) let $v_1,
\ldots, v_k$ be a path in $G^{(\rho)}$, 
then $var(v_1) < \ldots < var(v_k)$. We define the {\em height of a
node $v$ in a COBDD $\rho$} (notation ${\rm height}_\rho(v)$, or simply ${\rm
height}(v)$ if $\rho$ is understood) as the height of the DAG $G^{(\rho_v)}$, i.e.
the length of the longest path from $v$ to ${\bf 1}$ in $G^{(\rho)}$.

\subsection{Semantics of a COBDD}\label{semantics.subsec}

In Def.~\ref{flipped-obdd.semantics.def} we define the semantics
$\llbracket\cdot\rrbracket$ of each node $v \in V$ of a given COBDD $\rho =
({\cal V}$, $V$, ${\bf 1}$, ${\rm var}$, ${\rm low}$, ${\rm high}$, ${\rm
flip})$ as the boolean function represented by $v$, given the parity $b$ of
complemented edges seen on the path from a root to $v$. 

\begin{definition}\label{flipped-obdd.semantics.def}

Let $\rho = ({\cal V}$, $V$, ${\bf 1}$, ${\rm var}$, ${\rm low}$, ${\rm high}$,
${\rm flip})$ be a COBDD. The {\em semantics of a node $v \in V$ w.r.t. a 
flipping bit $b$} is a boolean function defined as:

\begin{itemize}

	\item $\llbracket {\bf 1}, b\rrbracket_\rho := \bar{b}$ (base of the
	induction)

	\item $\llbracket v, b\rrbracket_\rho := x_i\llbracket {\rm high}(v),
b\rrbracket_\rho + \bar{x}_i\llbracket {\rm low}(v), b \oplus{\rm
flip}(v)\rrbracket_\rho$ for any internal node $v$ (recursive step), being $x_i
= {\rm var}(v)$.

\end{itemize}

When $\rho$ is understood, we will write $\llbracket\cdot\rrbracket$ instead of
$\llbracket\cdot\rrbracket_\rho$.

\end{definition}

Note that the semantics of a node of a COBDD $\rho$ is a function of variables
in ${\cal V}$ and of an additional boolean variable $b$. Thus, on each node {\em
two} boolean functions on ${\cal V}$ are defined (one for each value of $b$). It
can be shown (Prop.~\ref{flipped-obdd.semantics.prop}) that such boolean
functions are complementary.

\begin{fact}\label{flipped-obdd.semantics.prop}

Let $\rho = ({\cal V}$, $V$, ${\bf 1}$, ${\rm var}$, ${\rm low}$, ${\rm high}$,
${\rm flip})$ be a COBDD, let $v \in V$ be a node and $b \in \B$ be a flipping
bit. Then $\llbracket v, b\rrbracket = \overline{\llbracket v,
\bar{b}\rrbracket}$.

\begin{proof}

The proof is by induction on $v$. As base of the induction, we have $\llbracket
{\bf 1}, b\rrbracket = \bar{b} = \bar{\bar{\bar{b}}} = \overline{\llbracket
{\bf 1}, \bar{b}\rrbracket}$.

As induction step, let $v$ be an internal node, and suppose by induction that
$\llbracket {\rm high}(v), b\rrbracket = \overline{\llbracket {\rm
high}(v), \bar{b}\rrbracket}$ and $\llbracket {\rm low}(v),
b\rrbracket = \overline{\llbracket {\rm low}(v), \bar{b}\rrbracket}$.

\sloppy

Then, since $AB + \bar{A}C = (\bar{A} + B) (A + C)$, we have: $\llbracket v,
b\rrbracket = x_i\llbracket {\rm high}(v), b\rrbracket +
\bar{x}_i\llbracket {\rm low}(v), b \oplus{\rm flip}(v)\rrbracket =$
$(\bar{x}_i + \llbracket {\rm high}(v), b\rrbracket) (x_i + \llbracket {\rm
low}(v), b \oplus{\rm flip}(v)\rrbracket) =$ $(\bar{x}_i +
\overline{\llbracket {\rm high}(v), \bar{b}\rrbracket}) (x_i +
\overline{\llbracket {\rm low}(v), \overline{b \oplus{\rm
flip}(v)}\rrbracket}) =$ $(\bar{x}_i + \overline{\llbracket {\rm high}(v),
\bar{b}\rrbracket}) (x_i + \overline{\llbracket {\rm low}(v), \bar{b}
\oplus{\rm flip(v)}\rrbracket}) =$ $\overline{x_i \llbracket {\rm high}(v),
\bar{b}\rrbracket + \bar{x}_i \llbracket {\rm low}(v), \bar{b} \oplus{\rm
flip(v)}\rrbracket} =$ $\overline{\llbracket v, \bar{b}\rrbracket}$.

\fussy

\end{proof}

\end{fact}

\begin{example}\label{cobdd-sem.ex}

Let $\rho = (\{x_0$, $x_1$, $x_2\}$, $\{{\rm 0x15}$, ${\rm 0x14}$, ${\rm 0x13}$,
${\rm 0xe}$, ${\bf 1}\}$, ${\bf 1}$, ${\rm var}$, ${\rm low}$, ${\rm high}$,
${\rm flip})$ be the COBDD of Ex.~\ref{cobdd.ex}. If we pick nodes ${\rm 0xe}$
and ${\rm 0x14}$ we have $\llbracket {\rm 0xe}, b\rrbracket = x_2\llbracket {\bf
1}, b\rrbracket + \bar{x}_2\llbracket {\bf 1}, b \oplus 1\rrbracket = x_2\bar{b}
+ \bar{x}_2b = x_2 \oplus b$ and $\llbracket {\rm 0x14}, b\rrbracket =
x_1\llbracket {\rm 0xe}, b\rrbracket + \bar{x}_1\llbracket {\bf 1}, b \oplus
0\rrbracket = x_1x_2\bar{b} + x_1\bar{x}_2b + \bar{x}_1\bar{b} = x_2\bar{b} +
x_1\bar{x}_2b + \bar{x}_1\bar{b}$.

Moreover, if we pick node ${\rm 0x14}$, then it represents the two following
boolean functions: $\llbracket {\rm 0x14}, 0\rrbracket = x_2 + \bar{x}_1$ and
$\llbracket {\rm 0x14}, 1\rrbracket = x_1\bar{x}_2$ (note that $\llbracket {\rm
0x14}, 0\rrbracket = \overline{\llbracket {\rm 0x14}, 1\rrbracket}$).

\end{example}

Theor.~\ref{canonical.theor} states that COBDDs are a {\em canonical} representation
for boolean functions (see~\cite{BRB90,MIY90}).

%

\begin{theorem}\label{canonical.theor}

Let $f : \B^n \to \B$ be a boolean function. Then there exist a COBDD $\rho =
({\cal V}$, $V$, ${\bf 1}$, ${\rm var}$, ${\rm low}$, ${\rm high}$, ${\rm
flip})$, a node $v \in V$ and a flipping bit $b \in \B$ s.t. $\llbracket v,
b\rrbracket = f(x)$. 
Moreover, let $\rho = ({\cal V}$, $V$, ${\bf 1}$, ${\rm var}$, ${\rm
low}$, ${\rm high}$, ${\rm flip})$ be a COBDD, let $v_1, v_2 \in V$ be nodes and
$b_1, b_2 \in \B$ be flipping bits. Then $\llbracket v_1, b_1 \rrbracket =
\llbracket v_2, b_2 \rrbracket$ iff $v_1 = v_2 \land b_1 = b_2$.

\end{theorem}

\sloppy

Efficient (i.e., at most $O(|V|\log|V|)$) algorithms~\cite{BRB90,MIY90} exist to
compute standard logical operations on COBDDs. We will assume to have available
the following functions (for instantiation and existential quantifier elimination): 

\begin{itemize}


	\item \fun{COBDD\_APP} s.t. $\langle v_{APP},$ $b_{APP}\rangle$ $=$
$\fun{COBDD\_APP}(x_{i_1},$ $\ldots,$ $x_{i_k},$ $v_{1},$ $b_1,$ $\ldots,$
$v_{k},$ $b_k,$ $v, b)$ iff $\llbracket v_{APP}, b_{APP}\rrbracket = \llbracket
v, b\rrbracket|_{x_{i_1} = \llbracket v_{1}, b_1\rrbracket, \ldots, x_{i_k} =
\llbracket v_{k}, b_k\rrbracket}$;

	\item \fun{COBDD\_EX} s.t. $\langle v_{EX},$ $b_{EX}\rangle =
\fun{COBDD\_EX}(x_{i_1},$ $\ldots,$ $x_{i_k},$ $v,$ $b)$ iff $\llbracket v_{EX},
b_{EX}\rrbracket = \exists x_{i_1},$ $\ldots,$ $x_{i_k}\; \llbracket v,
b\rrbracket$. 

\end{itemize}

\fussy

Note that the above defined functions may create new COBDD nodes. We assume that
such functions also  properly update $V$, ${\rm var}$, ${\rm low}$, ${\rm high}$,
${\rm flip}$ inside COBDD $\rho$ (${\bf 1}$ and ${\cal V}$ are not affected).

\section{\sloppy Automatic Synthesis of C Code from a COBDD}\label{algo.tex}

%

Let $K(x_1, \ldots, x_n, u_1, \ldots, u_r)$ be an mgo for a given control
problem. Let $\rho = ({\cal V}$, $V$, ${\bf 1}$, ${\rm var}$, ${\rm low}$, ${\rm
high}$, ${\rm flip})$ be a COBDD s.t. there exist $v \in V$, $b \in \B$ s.t.
$\llbracket v, b\rrbracket = K(x_1, \ldots, x_n, u_1, \ldots, u_r)$. Thus,
${\cal V} = {\cal X} \dotcup {\cal U} = \{x_1, \ldots, x_n\} \dotcup \{u_1,
\ldots, u_r\}$ (we denote with $\dotcup$ the disjoint union operator, thus
${\cal X} \cap {\cal U} = \varnothing$). We will call variables $x_i \in {\cal
X}$ as {\em state variables} and variables $u_j \in {\cal U}$ as {\em
action variables}. 

%
%
%
%
%

We want to solve the boolean functional equation problem introduced in
Sect.~\ref{problem.tex} targeting a {\em software} implementation. We do this by
using a COBDD representing all our boolean functions. This allows us to exploit
COBDD node sharing. This results in an improvement for the method
in~\cite{Tro98}, which targets a software implementation but which does not
exploit sharing. Finally, we also synthesize the software (i.e., C code)
implementation for $f_1, \ldots, f_r$, which is not considered in~\cite{Tro98}.
Given that $K$ is an mgo, this results in an {\em optimal control software} for
the starting LTS.



\subsection{Synthesis Algorithm: Overview}\label{algo.subsec}


Our method \fun{Synthesize} takes as input $\rho$, $v$ and  $b$ s.t. $\llbracket
v, b\rrbracket = K(\mybold{x}, \mybold{u})$. Then, it returns as output a C
function {\tt void K(int *x, int *u)} with the following property: if, before a
call to {\tt K}, $\forall i$ {\tt x[$i - 1$]}$= x_i$ holds (array indexes in C
language begin from $0$) with $\mybold{x} \in {\rm Dom}(K)$, and after the call
to {\tt K}, $\forall i$ {\tt u[$i - 1$]}$= u_i$ holds, then $K(\mybold{x},
\mybold{u}) = 1$. Moreover, the WCET of function {\tt K} is at most $O(nr)$.


Note that our method \fun{Synthesize} provides an effective {\em implementation}
of the mgo $K$, i.e. a C function which takes as input the current
state of the LTS and outputs the action to be taken. Thus, {\tt K} is indeed a
control software.

Function \fun{Synthesize} is organized in two phases:

\begin{enumerate}

\sloppy

	\item \label{action_bits.lbl} starting from $\rho$, $v$ and $b$ (thus
from $K(\mybold{x}, \mybold{u})$), we generate COBDD nodes $v_1, \ldots, v_r$
and flipping bits $b_1, \ldots, b_r$ for boolean functions $f_1, \ldots, f_r$
s.t. each $f_i = \llbracket v_i, b_i\rrbracket$ takes as input the state bit
vector $\mybold{x}$ and computes the $i$-th bit $u_i$ of an output action bit
vector $\mybold{u}$, where $K(\mybold{x}, \mybold{u}) = 1$, provided that
$\mybold{x} \in {\rm Dom}(K)$. This computation is carried out in function
\fun{SolveFunctionalEq};

	\item \label{c_code.lbl} $f_1, \ldots, f_r$ are translated inside
function {\tt void K(int *x, int *u)}. This step is performed by maintaining the
structure of the COBDD nodes representing $f_1, \ldots, f_r$. This allows us to
exploit COBDD node sharing in the generated software. This phase is
performed by function \fun{GenerateCCode}.

\fussy

\end{enumerate}


Thus function \fun{Synthesize} is organized as in Alg.~\ref{synthesize.alg}.
Correctness for function \fun{Synthesize} is proved by
Theor.~\ref{correctness.theor}.

\begin{algorithm}
  \caption[Translating COBDDs to a C function]{Translating COBDDs to a C function}
  \label{synthesize.alg}
  \begin{algorithmic}[1]
    \REQUIRE COBDD $\rho = ({\cal V}$, $V$, ${\bf 1}$, ${\rm var}$, ${\rm low}$, ${\rm high}$,
${\rm flip})$, node $v \in V$, boolean $b \in \B$
    \ENSURE {\fun{Synthesize}($\rho, v, b$)}:
    \STATE $\langle v_1, b_1, \ldots, v_r, b_r\rangle \gets$
    \fun{SolveFunctionalEq}($\rho, v, b$) {\em /* first phase */}
    \STATE \fun{GenerateCCode}($\rho, v_1, b_1, \ldots, v_r, b_r$) {\em /*
    second phase */}
 \end{algorithmic}
\end{algorithm}  

\subsection{Synthesis Algorithm: Solving Functional Equation (First Phase)}\label{algo_1.subsec}

In this phase, starting from $\rho$, $v$ and $b$ (thus from $\llbracket v,
b\rrbracket = K(\mybold{x}, \mybold{u})$), we compute the COBDD nodes $v_1,
\ldots, v_r$ and flipping bits $b_1, \ldots, b_r$ having the following
properties:

\begin{itemize}

	\item for all $i \in [r]$, $\llbracket v_i, b_i\rrbracket =
	f_i(\mybold{x})$ (thus each $f_i : \B^n \to \B$ does not depend on
	$\mybold{u}$);

	\item for all $\mybold{x} \in {\rm Dom}(K)$, $K(\mybold{x},
f_1(\mybold{x}), \ldots, f_r(\mybold{x})) = 1$.

\end{itemize}

\sloppy

In a hardware synthesis setting,  techniques to compute  $f_1, \ldots, f_r$ 
satisfying the above functional equation  have been widely studied (e.g.
see~\cite{BCK09}). In our software synthesis setting we follow an approach
similar to the one presented in~\cite{Tro98} to compute such $f_1, \ldots, f_r$.
Namely, we observe that $f_i$ may be computed using $f_1, \ldots, f_{i - 1}$,
that is $f_i(\mybold{x}) = \exists u_{i + 1},\ldots, u_n\; K(\mybold{x},
f_1(\mybold{x}), \ldots, f_{i - 1}(\mybold{x}), 1, u_{i + 1}, \ldots, u_n)$ (see
Lemma~\ref{functional.lmm}). This allows us to compute COBDD nodes $v_1, \ldots,
v_r$ and  flipping bits $b_1, \ldots, b_r$ as it is shown in function
\fun{SolveFunctionalEq} of Alg.~\ref{actions.alg}. Correctness for function
\fun{SolveFunctionalEq} is proved in Lemma~\ref{first_phase.lmm}.

\fussy

\begin{algorithm}
  \caption[Solving a Boolean Functional Equation]{Solving a boolean functional equation}
  \label{actions.alg}
  \begin{algorithmic}[1]
    \REQUIRE COBDD $\rho = ({\cal V}$, $V$, ${\bf 1}$, ${\rm var}$, ${\rm low}$, ${\rm high}$,
${\rm flip})$, node $v \in V$, boolean $b \in \B$
    \ENSURE {\fun{SolveFunctionalEq}($\rho, v, b$)}:
    \FORALL{$i \in [r]$}
      \STATE $\llbracket v_i,$ $b_i\rrbracket \gets \fun{COBDD\_EX}(u_{i + 1},
\ldots,$ $u_n,$ $\fun{COBDD\_APP}(u_1,$ $\ldots,$ $u_i,$ $v_{1},$ $b_1,$ $\ldots,$ $v_{i - 1},
b_{i - 1},$ ${\bf 1},$ $0,$ $v,$ $b))$
    \ENDFOR
    \RETURN $\langle v_1, b_1, \ldots, v_r, b_r\rangle$
 \end{algorithmic}
\end{algorithm}  

\subsection{Synthesis Algorithm: Generating C Code (Second Phase)}\label{algo_2.subsec}

In this phase, starting from COBDD nodes $v_1, \ldots, v_r$ and  flipping bits
$b_1, \ldots, b_r$ for functions $f_1, \ldots, f_r$ generated in the first
phase, we generate two C functions:

\begin{itemize} 

	\item {\tt void K(int *x, int *u)}, which is the required output
	function for our method \fun{Synthesize};

\sloppy

	\item {\tt int K\_bits(int *x, int action)}, which is an auxiliary
	function called by {\tt K}. A call to {\tt K\_bits(x, $i$)} returns
	$f_i(\mybold{x})$, being {\tt x[$j - 1$]}$= x_j$ for all $j \in [n]$.

\fussy

\end{itemize}

This phase is detailed in Algs.~\ref{translate_all.alg} and~\ref{translate.alg}.

\paragraph{Details of Function \fun{GenerateCCode} (Alg.~\ref{translate_all.alg})}\label{algo_2_1.subsec}

\begin{algorithm}
  \caption[Generating C functions]{Generating C functions}
  \label{translate_all.alg}
  \begin{algorithmic}[1]
    \REQUIRE COBDD $\rho = ({\cal V}$, $V$, ${\bf 1}$, ${\rm var}$, ${\rm low}$, ${\rm high}$,
${\rm flip})$, nodes $v_1, \ldots, v_r$, boolean values $b_1, \ldots, b_r$
    \ENSURE {\fun{GenerateCCode}($\rho, v_1, b_1, \ldots, v_r, b_r$)}:
    \PRINT ``{\tt int K\_bits(int *x, int action) \{ int ret\_b; switch(action)
    \{}''\label{begin_switch.algstep}
    \FORALL{$i \in [r]$}
      \PRINT ``{\tt case }'', $i - 1$, ``{\tt : ret\_b = }'', $\bar{b}_i$, ``{\tt
      ; goto L\_}'', $v_i$,``{\tt ;}''\label{select_bit_and_jump.algstep}
    \ENDFOR
    \PRINT ``{\tt \}}'' {\em /* end of the switch block */}\label{end_switch.algstep}
    \STATE $W \gets \varnothing$ \label{begin_nodes.algstep}
    \FORALL{$i \in [r]$}
      \STATE $W \gets$\fun{Translate}($\rho, v_i, W$);
      \label{call_translate.algstep} \label{end_nodes.algstep}
    \ENDFOR
    \PRINT ``{\tt \}}'' {\em /* end of K\_bits */}
    \PRINT ``{\tt void K(int *x,int *u)\{int i; for(i=0;i<}'', $r$, ``{\tt
    ;i++) u[i]=K\_bits(x,i);\}}''\label{printK.algstep}
 \end{algorithmic}
\end{algorithm}  

Given inputs $\rho,$ $v_1,$ $b_1,$ $\ldots,$ $v_r,$ $b_r$ (output by
\fun{SolveFunctionalEq}), Alg.~\ref{translate_all.alg} works as follows. First,
function {\tt int K\_bits(int *x, int action)} is generated. If {\tt x[$j -
1$]}$= x_j$ for all $j \in [n]$, the call {\tt K\_bits(x, $i$)} has to return
$f_i(\mybold{x})$. In order to do this, the graph $G^{(\rho_{v_i})}$ is
traversed by taking, in each node $v$, the then edge if ${\tt x[}j - 1{\tt ]} =
1$ (with $j$ s.t. ${\rm var}(v) = x_j$) and the else edge otherwise. When node
${\bf 1}$ is reached, then $1$ is returned iff the integer sum $c + b_i$ is
even, being $c$ the number of complemented else edges traversed. Note that
parity of $c + b_i$ may be maintained by initializing a C
variable {\tt ret\_b} to $\bar{b}_i$, then  complementing {\tt ret\_b}  (i.e.,
by performing a {\tt ret\_b = !ret\_b} statement) when a complemented else edge
is traversed, and finally returning {\tt ret\_b}. Note that formally this is
equivalent to compute the flipping bit $b$ s.t. $\langle {\bf 1}, \bar{b}\rangle
= \fun{COBDD\_APP}(x_1, \ldots, x_n, {\bf 1}, 1 - {\tt x[0]}, \ldots, {\bf 1}, 1
- {\tt x[}n - 1{\tt ]}, v_i, b_i)$, being $\llbracket v_i, b_i\rrbracket =
f_i(\mybold{x})$. 

This mechanism is implemented inside function {\tt K\_bits} by properly
translating each COBDD node $\tilde{v} \in \bigcup_{i = 1}^r V_{v_i}$  in a C
code block. Each block is labeled with a unique label depending on $\tilde{v}$,
and maintains in variable {\tt ret\_b} the current parity of $c + b_i$ as
described above. This is done by function \fun{Translate}, called on
line~\ref{call_translate.algstep} and detailed in Alg.~\ref{translate.alg}. 

Thus, the initial part of function {\tt K\_bits} consists of a {\tt switch}
block (generated in lines~\ref{begin_switch.algstep}--\ref{end_switch.algstep}
of Alg.~\ref{translate_all.alg}) which initializes {\tt ret\_b} to $\bar{b}_i$
and then jumps to the label corresponding to node $v_i$. Then, the C code blocks
corresponding to COBDD nodes are generated in
lines~\ref{begin_nodes.algstep}--\ref{end_nodes.algstep} of
Alg.~\ref{translate_all.alg}, by calling $r$ times function \fun{Translate} (see
Alg.~\ref{translate.alg}) with parameters $v_1, \ldots, v_r$. Note that $W$
maintains the already translated COBDD nodes. Since function \fun{Translate}
only translates nodes not in $W$, this allows us to exploit sharing not only
inside each $G^{(\rho_{v_i})}$, but also inside $G^{(\rho_{v_1})}, \ldots,
G^{(\rho_{v_r})}$.

Finally, function {\tt K} is generated in line~\ref{printK.algstep}. Function
{\tt K} simply consists in a {\tt for} loop filling each entry {\tt u[i]} of the
output array {\tt u} with the boolean values returned by {\tt K\_bits(x, i)}.
Correctness of function \fun{GenerateCCode} is proved in
Lemma~\ref{translate_corr.lmm}.

\paragraph{Details of Function \fun{Translate} (Alg.~\ref{translate.alg})}\label{algo_2_2.subsec}

\begin{algorithm}
  \caption[COBDD Nodes Translation]{COBDD nodes translation}
  \label{translate.alg}
  \begin{algorithmic}[1]
    \REQUIRE COBDD $\rho = ({\cal V}$, $V$, ${\bf 1}$, ${\rm var}$, ${\rm low}$, ${\rm high}$,
${\rm flip})$, node $v$, nodes set $W
\subseteq V$
    \ENSURE {\fun{Translate}($\rho, v, W$)}:
    \STATE \colorbox{light-gray}{{\bf if} $v \in W$ {\bf then return $W$}} \label{skip_visited_check.algstep}\label{skip_visited.algstep}
    \STATE $W \gets W \cup \{v\}$\label{updateW.algstep}
    \PRINT ``{\tt L\_}'', $v$, ``{\tt :}''\label{begin_c_code.algstep}
    \IF {$v$ $=$ ${\bf 1}$}\label{terminal_node_check.algstep}
      \PRINT ``{\tt return ret\_b;}''\label{terminal_node.algstep}
    \ELSE
      \STATE let $i$ be s.t. ${\rm var}(v) = x_i$ \label{begin_internal.algstep}
      \PRINT ``{\tt if (x[}'', $i - 1$, ``{\tt ] == 1) goto L\_}'',
      ${\rm high}(v)$, ``{\tt ;}''\label{then_edge.algstep}
      \STATE {\bf if} ${\rm flip}(v)$ {\bf then print} ``{\tt else \{ret\_b = !ret\_b;goto L\_}'',${\rm
	low}(v)$,``{\tt ;\}}''\label{compl_else_edge.algstep} \label{check_flip.algstep}
      \STATE {\bf else print} ``{\tt else goto L\_}'', ${\rm
	low}(v)$,``{\tt	;}''\label{else_edge.algstep}\label{end_c_code.algstep}\label{end_internal.algstep}
      \STATE $W \gets$\fun{Translate}($\rho, {\rm high}(v), W$)
      \label{then_recursion.algstep}
      \STATE $W \gets$\fun{Translate}($\rho, {\rm low}(v), W$) \label{else_recursion.algstep}
    \ENDIF
    \RETURN $W$
 \end{algorithmic}
\end{algorithm}  

Given inputs $\rho, v, W$, Alg.~\ref{translate.alg} performs a recursive graph
traversal of $G^{(\rho_v)}$ as follows. 

The C code block for internal node $v$ is generated in
lines~\ref{begin_c_code.algstep}
and~\ref{begin_internal.algstep}--\ref{end_internal.algstep}. 
The block consists of a label {\tt L\_$v$:} and an {\tt if-then-else} C
construct. Note that label {\tt L\_$v$} univocally identifies the C code block
related to node $v$. This may be implemented by printing the exadecimal value of
a pointer to $v$.

The {\tt if-then-else} C construct is generated so as to traverse node $v$ in
graph $G^{(\rho_v)}$ in the following way. In line~\ref{then_edge.algstep} the
check {\tt x[$i - 1$]}$ = 1$ is generated, being $i$ s.t. ${\rm var}(v) = x_i$.
The code to take the then edge of $v$ is also generated. Namely, it is
sufficient to generate a {\tt goto} statement to the C code block related to
node ${\rm high}(v)$. In lines~\ref{compl_else_edge.algstep}
and~\ref{else_edge.algstep} the code to take the else edge is generated, in the
case {\tt x[$i - 1$]}$ = 1$ is false. In this case, if the else edge is
complemented, i.e. ${\rm flip}(v)$ holds (line~\ref{check_flip.algstep}), it is
necessary to complement {\tt ret\_b} and then perform a {\tt goto} statement to
the C code block related to node ${\rm low}(v)$
(line~\ref{compl_else_edge.algstep}). Otherwise, it is sufficient to generate a
{\tt goto} statement to the C code block related to node ${\rm low}(v)$
(line~\ref{else_edge.algstep}). 

Thus, the block generated for an internal node $v$, for proper $i$, $l$ and $h$,
has one of the following forms:

\begin{itemize}

	\item {\tt L\_$v$: if (x[$i - 1$]) goto L\_$h$; else goto L\_$l$;} 

	\item {\tt L\_$v$: if (x[$i - 1$]) goto L\_$h$; else \{ret\_b = !ret\_b;
goto L\_$l$;\}}.

\end{itemize}

There are two base cases for the recursion of function \fun{Translate}:

\begin{itemize}

	\item $v \in W$ (line~\ref{skip_visited_check.algstep}), i.e. $v$ has
already been translated into a C code block as above. In this case, the set of
visited COBDD nodes $W$ is directly returned (line~\ref{skip_visited.algstep})
without generating any C code. This allows us to retain COBDD node
sharing;

	\item $v = {\bf 1}$ (line~\ref{terminal_node_check.algstep}), i.e.
the terminal node ${\bf 1}$ has been reached. In this case, the C code block to
be generated is simply {\tt L\_1: return ret\_b;}. Note that such a
block will be generated only once.

\end{itemize}

In all other cases, function \fun{Translate} ends with the recursive calls on
the then and else edges
(lines~\ref{then_recursion.algstep}--\ref{else_recursion.algstep}). Note that
the visited nodes set $W$ passed to the second recursive call is the result of
the first recursive call. Correctness of function \fun{Translate} is proved in
Lemma~\ref{translate_corr.lmm}.

%

\begin{figure}
\begin{tabular}{ccc}
\begin{minipage}{0.4\textwidth}
\hspace*{-1cm}
\includegraphics[scale=0.4]{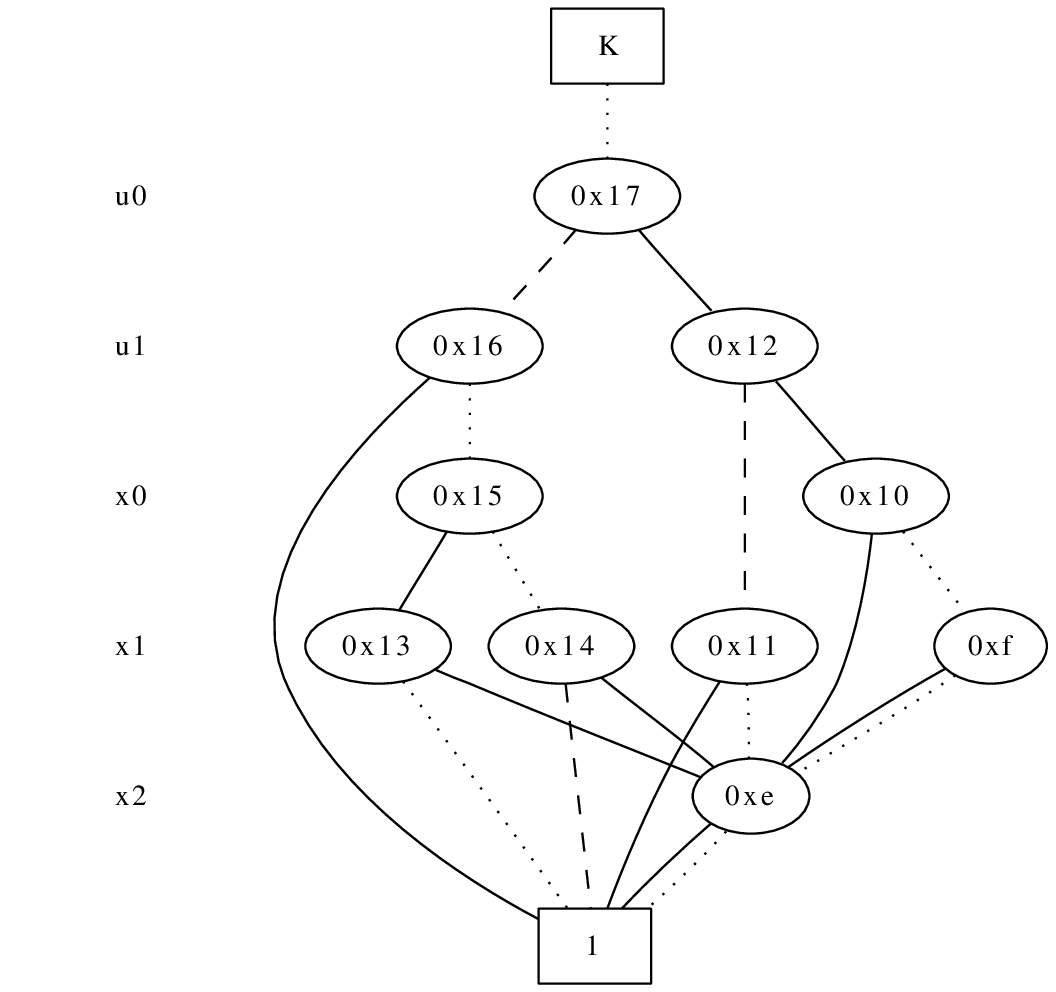}
\caption{An mgo example}
\label{K.example.eps}
\end{minipage}
&
\begin{minipage}{0.25\textwidth}
\includegraphics[scale=0.4]{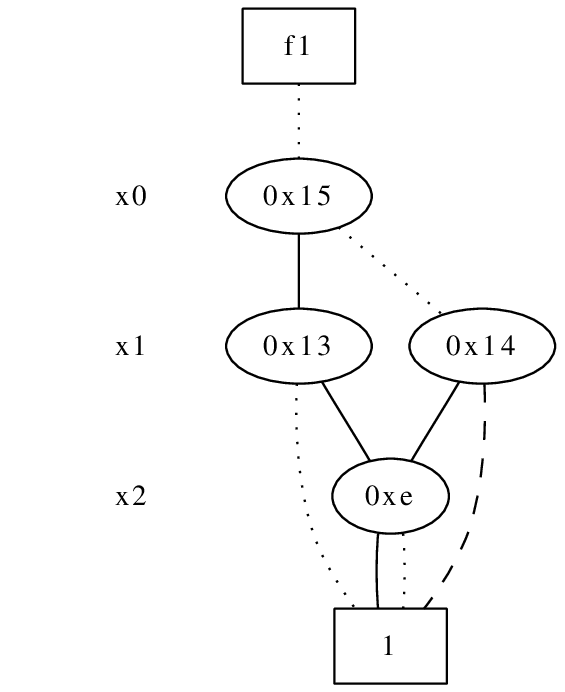}
\caption{Computing first action bit for mgo in Fig.~\ref{K.example.eps}}
\label{F1.example.eps}
\end{minipage}
&
\begin{minipage}{0.25\textwidth}
\includegraphics[scale=0.4]{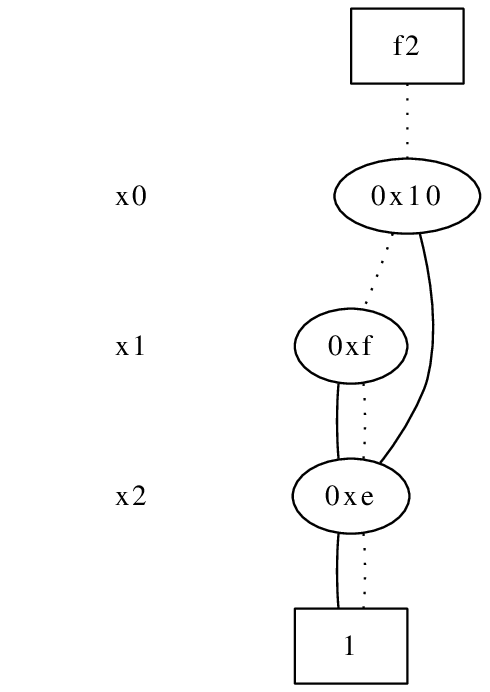}
\caption{Computing second action bit for mgo in Fig.~\ref{K.example.eps}}
\label{F2.example.eps}
\end{minipage}
\end{tabular}
\end{figure}

\subsection{An Example of Translation}\label{example.tex}

In this section we show how a node $v$ and a flipping bit $b$ of a COBDD
$\rho$ with 3 state variables and 2 action variables is translated in {\tt K}
and {\tt K\_bits} C functions. This is done by applying 
Algs.~\ref{synthesize.alg},~\ref{actions.alg},~\ref{translate_all.alg}
and~\ref{translate.alg}.

\sloppy

Consider COBDD $\rho = (\{u_0, u_1, x_0, x_1, x_2\}$, $\{{\rm 0x17}$, ${\rm
0x16}$, ${\rm 0x15}$, ${\rm 0x14}$, ${\rm 0x13}$, ${\rm 0x12}$, ${\rm 0x11}$, 
${\rm 0x10}$, ${\rm 0xf}$, ${\rm 0xe}$, ${\bf 1}\}$, ${\bf 1}$, ${\rm var}$,
${\rm low}$, ${\rm high}$, ${\rm flip})$. The corresponding $G^{(\rho)}$ is
shown in Fig.~\ref{K.example.eps}. Within $\rho$, consider mgo $K(x_0, x_1, x_2,
u_0, u_1) = \llbracket {\rm 0x17}, 1\rrbracket =
\bar{u}_0\bar{u}_1\bar{x}_0x_1\bar{x}_2 + \bar{u}_0\bar{u}_1x_0x_1x_2 +
u_0\bar{u}_1\bar{x}_1x_2 + u_0u_1\bar{x}_0\bar{x}_1\bar{x}_2 +
u_0u_1\bar{x}_0x_1x_2 + u_0u_1x_0\bar{x}_2$.
%
%
%
%
%
%
By applying \fun{SolveFunctionalEq} (see Alg.~\ref{actions.alg}), we obtain $f_1(x_0,
x_1, x_2) = \llbracket {\rm 0x15}, 1\rrbracket = \bar{x}_0\bar{x}_1 +
\bar{x}_0x_1x_2 + x_0\bar{x}_1 + x_0x_1\bar{x}_2$ and  $f_2(x_0, x_1, x_2) =
\llbracket {\rm 0x10}, 1\rrbracket = \bar{x}_0\bar{x}_1\bar{x}_2 +
\bar{x}_0x_1x_2 + x_0\bar{x}_2$. COBDDs for $f_1$ and $f_2$ are depicted in
Figs.~\ref{F1.example.eps} and~\ref{F2.example.eps} respectively. Note that in
this simple example no
new nodes have been added w.r.t. the COBDD of Fig.~\ref{K.example.eps}, and that
node ${\rm 0xe}$ is shared between $G^{(\rho_{\rm 0x15})}$ and
$G^{(\rho_{\rm 0x10})}$.
%
%
%
Finally, by calling \fun{GenerateCCode} (see Alg.~\ref{translate_all.alg}) on $f_1, f_2$, we have the C
code in Fig.~\ref{ctrl.x3.u2.c.tex}.

\fussy

\begin{figure}
  \framebox[1.0\hsize][c]{
    \begin{minipage}{0.9\hsize}
      \begin{center}
        \small
        \input{ctrl.x3.u2.c.tex}
      \end{center}
    \end{minipage}
  }
  \caption{C code for mgo in Fig.~\ref{K.example.eps}} \label{ctrl.x3.u2.c.tex}
\end{figure}

\section{Translation Proof of Correctness}\label{proof.tex}

In this section we prove the correctness of our approach
(Theor.~\ref{correctness.theor}). That is, we show that the function {\tt K} we
generate indeed implements the given mgo $K$, thus resulting in a
correct-by-construction control software. 

We begin by stating four useful lemmata for our proof.
Lemma~\ref{functional.lmm} is useful to prove Lemma~\ref{first_phase.lmm}, i.e.
to prove correctness of function \fun{SolveFunctionalEq}.

\sloppy

\begin{lemma}\label{functional.lmm}

Let $K : \B^n \times \B^r \to \B$ and let $f_1, \ldots, f_r$ be s.t.
$f_i(\mybold{x}) = \exists u_{i + 1},\ldots, u_r$ $K(\mybold{x},
f_1(\mybold{x}), \ldots, f_{i - 1}(\mybold{x}), 1, u_{i + 1}, \ldots, u_r)$ for
all $i \in [r]$. Then, $\mybold{x} \in {\rm Dom}(K)$ $\Rightarrow$
$K(\mybold{x}, f_1(\mybold{x}), \ldots, f_r(\mybold{x})) = 1$.

\end{lemma}

\begin{proof}

Let $\mybold{x} \in \B^n$ be s.t. $\mybold{x} \in {\rm Dom}(K)$, i.e. $\exists
\mybold{u}\; K(\mybold{x}, \mybold{u}) = 1$.  We prove the lemma by induction on
$r$. For $r = 1$, we have $f_1(\mybold{x}) = K(\mybold{x}, 1)$. If
$f_1(\mybold{x}) = 1$, we have $K(\mybold{x}, f_1(\mybold{x})) = K(\mybold{x},
1) = f_1(\mybold{x}) = 1$. If $f_1(\mybold{x}) = 0$, we have $K(\mybold{x},
f_1(\mybold{x})) = K(\mybold{x}, 0)$, and $K(\mybold{x}, 0) = 1$ since
$\mybold{x} \in {\rm Dom}(K)$ and $K(\mybold{x}, 1) = 0$. 

Suppose by induction that for all $\tilde{K} : \B^n \times \B^{r - 1} \to \B$
$\tilde{K}(x, \tilde{f}_1(\mybold{x}), \ldots, \tilde{f}_{r - 1}(\mybold{x})) =
1$, where for all $i \in [r - 1]$ $\tilde{f}_i(\mybold{x}) = \exists u_{i +
1},\ldots, u_{r - 1}\; \tilde{K}(\mybold{x}, \tilde{f}_1(\mybold{x}), \ldots,
\tilde{f}_{i - 1}(\mybold{x}), 1, u_{i + 1}, \ldots, u_{r - 1})$. 
We have that $\mybold{x} \in {\rm Dom}(K)$ implies that either $\mybold{x} \in
{\rm Dom}(K|_{u_1 = 0})$ or $\mybold{x} \in {\rm Dom}(K|_{u_1 = 1})$. Suppose
$\mybold{x} \in {\rm Dom}(K|_{u_1 = 1})$ holds. We have that $K|_{u_1 =
1}(\mybold{x}, \tilde{f}_2(\mybold{x}), \ldots, \tilde{f}_r(\mybold{x})) = 1$,
where for all $i = 2, \ldots, r$ $\tilde{f}_i(\mybold{x}) = \exists u_{i +
1},\ldots, u_r\; K|_{u_1 = 1}(\mybold{x}, \tilde{f}_2(\mybold{x}), \ldots,
\tilde{f}_{i - 1}(\mybold{x}), 1, u_{i + 1}, \ldots, u_r)$. By construction, we
have that $f_1(\mybold{x}) = 1$ and $f_i(\mybold{x}) = \tilde{f}_i(\mybold{x})$
for $i \geq 2$, thus $1 = K|_{u_1 = 1}(\mybold{x}, \tilde{f}_2(\mybold{x}),
\ldots, \tilde{f}_r(\mybold{x})) = K(\mybold{x}, f_1(\mybold{x}), \ldots,
f_r(\mybold{x}))$. Analogously, if $x \notin {\rm Dom}(K|_{u_1 = 1}) \land
\mybold{x} \in {\rm Dom}(K|_{u_1 = 0})$ we have that $f_1(\mybold{x}) = 0$ and
$f_i(\mybold{x}) = \tilde{f}_i(\mybold{x})$ for $i \geq 2$, thus $1 = K|_{u_1 =
0}(\mybold{x}, \tilde{f}_2(\mybold{x}), \ldots, \tilde{f}_r(\mybold{x})) =
K(\mybold{x}, f_1(\mybold{x}), \ldots, f_r(\mybold{x}))$.

\end{proof}

\fussy


Lemma~\ref{first_phase.lmm} states correctness of function \fun{SolveFunctionalEq} of
Alg.~\ref{actions.alg}. 

\begin{lemma}\label{first_phase.lmm}

Let $\rho = ({\cal V}$, $V$, ${\bf 1}$, ${\rm var}$, ${\rm low}$, ${\rm high}$,
${\rm flip})$ be a COBDD with ${\cal V} = {\cal X} \dotcup {\cal U}$, $v \in V$
be a node, $b \in \B$ be a flipping bit. Let $\llbracket v, b\rrbracket =
K(\mybold{x}, \mybold{u})$ and $r = |{\cal U}|$. Then function
\fun{SolveFunctionalEq}$(\rho, v, b)$ (see Alg.~\ref{actions.alg}) outputs nodes
$v_1, \ldots, v_r$ and boolean values $b_1, \ldots, b_r$ s.t. for all $i \in
[r]$ $\llbracket v_i, b_i \rrbracket = f_i(\mybold{x})$ and $\mybold{x} \in {\rm
Dom}(K)$ implies $K(\mybold{x}, f_1(\mybold{x}), \ldots, f_r(\mybold{x})) = 1$.

\end{lemma}

\begin{proof}


\sloppy

Correctness of functions \fun{COBDD\_APP} and \fun{COBDD\_EX} (and lemma
hypotheses) implies that for all $i \in [r]$ $f_i(\mybold{x}) = \exists u_{i +
1},\ldots, u_r\; K(\mybold{x}, f_1(\mybold{x}), \ldots, f_{i - 1}(\mybold{x}),
1, u_{i + 1}, \ldots, u_r)$. By Lemma~\ref{functional.lmm} we have the thesis.

\fussy

\end{proof}


Let \fun{Translate\_dup} be a function that works as function \fun{Translate} of
Alg.~\ref{translate.alg}, but that does not take node sharing into account.
Function \fun{Translate\_dup} may be obtained from function \fun{Translate} by
deleting line~\ref{skip_visited.algstep} (highlighted in
Alg.~\ref{translate.alg}) and by replacing calls to \fun{Translate} in
lines~\ref{then_recursion.algstep} and~\ref{else_recursion.algstep} with
recursive calls to \fun{Translate\_dup} (with no changes on parameters).
Lemma~\ref{translate_corr_noW.lmm} states correctness of function
\fun{Translate\_dup}.

\begin{lemma}\label{translate_corr_noW.lmm}

Let $\rho = ({\cal V}$, $V$, ${\bf 1}$, ${\rm var}$, ${\rm low}$, ${\rm high}$,
${\rm flip})$ be a COBDD, $v \in V$ be a node, $b \in \B$ be a flipping bit, and
$W \subseteq V$ be a  set of nodes. Then function \fun{Translate\_dup}$(\rho, v,
W)$ generates a sequence of labeled C statements $B_1 \ldots B_k$ s.t. $k \geq
|V_v|$ and for all $w \in V_v$: 1) label {\tt L\_$w$} is in $B_i$ for some $i$
and 2) starting an execution from label {\tt L\_$w$} with $\forall i \in
[n]$ {\tt x[$i - 1$]}$= x_i$ and {\tt ret\_b}$= \bar{b}$, a {\tt return ret\_b;}
statement is invoked in at most $O(p)$ steps with ${\tt ret\_b} = \llbracket w,
b\rrbracket = f_{w, b}(\mybold{x})$ and $p = {\rm height}(w)$.

\end{lemma}

\begin{proof}

\sloppy

We prove this lemma by induction on $v$. Let $v = {\bf 1}$, which implies
$\llbracket v, b\rrbracket = \bar{b}$ and $V_v = \{{\bf 1}\}$. We have that
function \fun{Translate\_dup}$(\rho, v, W)$ generates a single block $B_1$ (thus
$k = 1 = |V_{\bf 1}|$) s.t. $B_1 = ${\tt L\_1: return ret\_b;}
(lines~\ref{begin_c_code.algstep}--\ref{terminal_node.algstep} of
Alg.~\ref{translate.alg}). Since by hypothesis we have {\tt ret\_b}$= \bar{b}$,
and since starting from $B_1$ the return statement is invoked in $O(1)$ steps,
the base case of the induction is proved.

Let $v$ be an internal node with ${\rm var}(v) = x_i$ and let $f(\mybold{x}) = \llbracket
v, b\rrbracket$. Since $w \in V_v$ iff $w = v \lor w \in V_{{\rm high}(v)} \lor
w \in V_{{\rm low}(v)}$, by induction hypothesis we only have to prove the
thesis for $w = v$. We have that $f(\mybold{x}) = x_i\llbracket {\rm high}(v),
b\rrbracket + \bar{x}_i\llbracket {\rm low}(v), b \oplus{\rm
flip}(v)\rrbracket$, i.e. $f(\mybold{x}) = x_i\llbracket {\rm high}(v), b\rrbracket +
\bar{x}_i\llbracket {\rm low}(v), b\rrbracket$ if ${\rm flip}(v) = 0$ and $f(\mybold{x})
= x_i\llbracket {\rm high}(v), b\rrbracket + \bar{x}_i\llbracket {\rm low}(v),
\bar{b}\rrbracket$ if ${\rm flip}(v) = 1$. Since $f(\mybold{x}) = x_if|_{x_i = 1}(\mybold{x}) +
\bar{x}_if|_{x_i = 0}(\mybold{x})$, by Theor.~\ref{canonical.theor} we have that
$\llbracket {\rm high}(v), b\rrbracket =  f|_{x_i = 1}(\mybold{x})$, and that $\llbracket
{\rm low}(v), b\rrbracket =  f|_{x_i = 0}(\mybold{x})$ if ${\rm flip}(v) = 0$ and
$\llbracket {\rm low}(v), \bar{b}\rrbracket =  f|_{x_i = 0}(\mybold{x})$ if ${\rm
flip}(v) = 1$. 

By lines~\ref{begin_c_code.algstep}
and~\ref{then_edge.algstep}--\ref{else_edge.algstep} of
Alg.~\ref{translate.alg}, we have that function \fun{Translate\_dup}$(\rho, v,
W)$ generates blocks $B B_{11} \ldots B_{1h} B_{21} \ldots B_{2l}$ s.t. $B =
${\tt L\_$v$: if (x[$i - 1$] == 1) goto L\_${\rm high}(v)$; else }$B_E$ where
$B_E$ is either {\tt goto L\_${\rm low}(v)$;} if ${\rm flip}(v) = 0$ or {\tt
\{ret\_b = !ret\_b; goto L\_${\rm low}(v)$;\}} if ${\rm flip}(v) = 1$, and
$B_{11} \ldots B_{1h}$ ($B_{21} \ldots B_{2l}$) are generated by the recursive
call \fun{Translate\_dup}$(\rho, {\rm high}(v), W)$ in
line~\ref{then_recursion.algstep} (\fun{Translate\_dup}$(\rho, {\rm low}(v), W)$
in line~\ref{else_recursion.algstep}). By induction hypothesis and the above
reasoning, if the execution starts at label {\tt L\_${\rm high}(v)$} and {\tt
ret\_b}$= \bar{b}$, then a {\tt return ret\_b;} statement is invoked in at most
$O(p - 1)$ steps with ${\tt ret\_b} = f|_{x_i = 1}(\mybold{x})$. As for the else
case, we have that starting from {\tt L\_${\rm low}(v)$} with {\tt ret\_b}$=
\bar{b}$ ({\tt ret\_b}$= \bar{\bar{b}}$) if ${\rm flip}(v) = 0$ (${\rm flip}(v)
= 1$), then a {\tt return ret\_b;} statement is invoked in at most $O(p - 1)$
steps with ${\tt ret\_b} = f|_{x_i = 0}(\mybold{x})$. By construction of block
$B$, starting from label {\tt L\_$v$}, a {\tt return ret\_b;} statement is
invoked in at most $O(p - 1 + 1) = O(p)$ steps with ${\tt ret\_b} = x_if|_{x_i =
1}(\mybold{x}) + \bar{x}_if|_{x_i = 0}(\mybold{x}) = f(\mybold{x})$. Finally,
note that by induction hypothesis $h \geq |V_{{\rm high}(v)}|$ and $l \geq
|V_{{\rm low}(v)}|$, thus we have that $k = 1 + h + l \geq 1 + |V_{{\rm
high}(v)}| + |V_{{\rm low}(v)}| \geq |V_v|$.  

\fussy

\end{proof}

Lemma~\ref{translate_corr.lmm} extends Lemma~\ref{translate_corr_noW.lmm} by
also considering node sharing, thus stating correctness of function
\fun{GenerateCCode} of Alg.~\ref{translate_all.alg} and function \fun{Translate}
of Alg.~\ref{translate.alg}.

\begin{lemma}\label{translate_corr.lmm}

Let $\rho = ({\cal V}$, $V$, ${\bf 1}$, ${\rm var}$, ${\rm low}$, ${\rm high}$,
${\rm flip})$ be a COBDD and $v_1, \ldots, v_r \in V$ be $r$ nodes and $b_1,
\ldots, b_r \in \B$ be $r$ flipping bits. Then
lines~\ref{begin_nodes.algstep}--\ref{end_nodes.algstep} of function
\fun{GenerateCCode}$(\rho, v_1, b_1, \ldots, v_r, b_r)$ generate a sequence of
labeled C statements $B_1 \ldots B_k$ s.t. $k = |\cup_{i=1}^r V_{v_i}|$ and for
all $v \in \cup_{i=1}^r V_{v_i}$: 1) the label {\tt L\_$v$} is in $B_j$ for some
$j$ and 2) starting an execution from label {\tt L\_$v$} with $\forall j \in [n]$
{\tt x[$j - 1$]}$= x_j$ and {\tt ret\_b}$= \bar{b}$, a {\tt return ret\_b;}
statement is invoked in at most $O(p)$ steps with ${\tt ret\_b} = \llbracket v,
b\rrbracket = f_{v, b}(\mybold{x})$ and $p = {\rm height}(w)$.

%

\end{lemma}

\begin{proof}

We begin by proving that $k = |\cup_{i=1}^r V_{v_i}|$. To this aim, we prove
that for each node $v \in \cup_{i=1}^r V_{v_i}$, a unique block $B_v$ is
generated. This follows by how the nodes set $W$ is managed by function
\fun{Translate} in lines~\ref{skip_visited.algstep}--\ref{begin_c_code.algstep}
of Alg.~\ref{translate.alg} and by function \fun{GenerateCCode} in
lines~\ref{begin_nodes.algstep}--\ref{end_nodes.algstep} of
Alg.~\ref{translate_all.alg}. In fact, function \fun{Translate}, when called on
parameters $\rho, v, W$, returns a set $W' \supseteq W$, and function
\fun{GenerateCCode} calls \fun{Translate} by always passing the $W$ resulting by
the previous call. Since a block is generated for node $v$ only if $v$ is not in
$W$, and $v$ is added to $W$ only when a block is generated for node $v$, this
proves this part of the lemma.

As for correctness, we prove this lemma by induction on $m$, being $m$ the
number of times that the {\tt return $W$;} statement in
line~\ref{skip_visited.algstep} of Alg.~\ref{translate.alg} is executed. As base
of the induction, let $m = 1$ and let $\rho, v, W$ be the parameters of the
recursive call executing the first {\tt return $W$;} statement. Then, by
construction of function \fun{Translate}, $v$ has been added to $W$ in some
previous recursive call with parameters $\rho, v, \tilde{W}$. In this previous
recursive call, a block $B_v$ with label {\tt L\_$v$} has been generated.
Moreover, for this previous recursive call, thus for parameters $\rho, v,
\tilde{W}$, we are in the hypothesis of Lemma~\ref{translate_corr_noW.lmm},
which implies that the induction base is proved.

Suppose now that the thesis holds for the first $m$ executions of the {\tt
return $W$;} statement in line~\ref{skip_visited.algstep} of
Alg.~\ref{translate.alg}. Then, by construction of function \fun{Translate}, $v$
has been added to $W$ in some previous recursive call with parameters $\rho, v,
\tilde{W}$. In this previous recursive call, a block $B_v$ with label {\tt
L\_$v$} has been generated. Let $w_1, W_1, \ldots, w_m, W_m$, be s.t. the $m$
recursive calls executing the {\tt return $W$;} statement have parameters $\rho,
v_i, W_i$ (note that they are not necessarily distinct). By induction
hypothesis, for all $i \in [m]$ starting from label {\tt L\_$w_i$} with $\forall
j \in [n]$ {\tt x[$j - 1$]}$= x_j$ and {\tt ret\_b}$= \bar{b}$, a {\tt return
ret\_b;} statement is invoked in at most $O(p)$ steps with ${\tt ret\_b} =
f_{w_i, b}(\mybold{x})$. By Lemma~\ref{translate_corr_noW.lmm} and its proof, the same
holds for all $v \in V_v \setminus \{w_1, \ldots, w_m\}$, thus it holds for all
$v \in V_v$. 

\end{proof}

We are now ready to give our main correctness theorem for function
\fun{Synthesize} of Alg.~\ref{synthesize.alg}.

\begin{theorem}\label{correctness.theor}

Let $\rho = ({\cal V}$, $V$, ${\bf 1}$, ${\rm var}$, ${\rm low}$, ${\rm high}$,
${\rm flip})$ be a COBDD with ${\cal V} = {\cal X} \dotcup {\cal U}$, $v \in V$
be a node, $b \in \B$ be a boolean. Let $\llbracket v, b\rrbracket =
K(\mybold{x}, \mybold{u})$, $r = |{\cal U}|$ and $n = |{\cal X}|$. Then function
\fun{Synthesize}$(\rho, v, b)$ generates a C function {\tt void K(int *x, int
*u)} with the following property: for all $\mybold{x} \in {\rm Dom}(K)$, if
before a call to {\tt K} $\forall i \in [n]$ {\tt x[$i - 1$]}$= x_i$, and after
the call to {\tt K} $\forall i \in [r]$ {\tt u[$i - 1$]}$= u_i$, then
$K(\mybold{x}, \mybold{u}) = 1$. 

Furthermore, function {\tt K} has WCET $\sum_{i = 1}^{r}O({\rm height}(v_i))$,
being $v_1, \ldots, v_r$ the nodes output by function \fun{SolveFunctionalEq}. 

\end{theorem}

\begin{proof}

Let $\mybold{x} \in {\rm Dom}(K)$ (i.e. $\exists \mybold{u} \; K(\mybold{x},
\mybold{u}) = 1$) and suppose that for all $j \in [n]$ {\tt x[$j - 1$]}$= x_j$.
By line~\ref{printK.algstep} of Alg.~\ref{translate_all.alg}, for all $i \in
[r]$, {\tt u[$i - 1$]} will take the value returned by {\tt K\_bits(x, $i$)}. In
turn, by line~\ref{select_bit_and_jump.algstep} Alg.~\ref{translate_all.alg},
each {\tt K\_bits(x, $i$)} sets {\tt ret\_b} to $\bar{b}_i$ and makes a jump to
label {\tt L\_$v_i$}. By Lemma~\ref{first_phase.lmm} and by construction of
\fun{Synthesize}, such $b_1, \ldots, b_r$ and $v_1, \ldots, v_r$ are s.t. that
$\llbracket v_1, b_1\rrbracket = f_1(\mybold{x}), \ldots, \llbracket v_r,
b_r\rrbracket = f_r(\mybold{x})$ and $K(\mybold{x}, f_1(\mybold{x}), \ldots,
f_r(\mybold{x})) = 1$. By Lemma~\ref{translate_corr.lmm}, the sequence of calls
{\tt K\_bits(x, $1$)}, \ldots, {\tt K\_bits(x, $r$)} will indeed return, in at
most $\sum_{j = 1}^r O({\rm height}(v_i))$ steps, $f_1(\mybold{x}), \ldots,
f_r(\mybold{x})$.  

%

\end{proof}

\begin{corollary}\label{wcet.cor}

Let $\rho = ({\cal V}$, $V$, ${\bf 1}$, ${\rm var}$, ${\rm low}$, ${\rm high}$,
${\rm flip})$ be a COBDD with ${\cal V} = {\cal X} \dotcup {\cal U}$, $v \in V$
be a node, $b \in \B$ be a boolean. Let $\llbracket v, b\rrbracket =
K(\mybold{x}, \mybold{u})$, $r = |{\cal U}|$ and $n = |{\cal X}|$. Then the C
function {\tt K} output by function \fun{Synthesize}$(\rho, v, b)$ has WCET
$O(rn)$. 

\end{corollary}

\begin{proof}

The corollary immediately follows from Theor.~\ref{correctness.theor} and from
the fact that, for all $v \in V$, ${\rm height}(v) \leq n$.

\end{proof}

\section{Experimental Results}\label{expres.tex}

We implemented our synthesis algorithm  in C programming language, using the
CUDD package for OBDD based computations. We name the resulting tool KSS ({\em
Kontrol Software Synthesizer}).  KSS is  part of a more general tool named QKS
({\em Quantized feedback Kontrol Synthesizer}~\cite{qks-cav2010}). KSS takes as
input a BLIF file which encodes the OBDD for an mgo $K(\mybold{x}, \mybold{u})$.
Such BLIF file also contains information about how to distinguish from state
variables $\mybold{x}$ and action variables $\mybold{u}$. Then KSS generates as
output a C code file containing functions {\tt K} and {\tt K\_bits} as described
in Sect.~\ref{algo.tex}. In this section we present our experiments that aim at
evaluating effectiveness of KSS.

\subsection{Experimental Settings} \label{expres_setting.subsec}

We present experimental results obtained by using KSS on given COBDDs $\rho_1,
\ldots, \rho_4$ s.t. for all $i \in [4]$:

\begin{itemize}

\sloppy

	\item $\rho_i = ({\cal V}_i$, $V_i$, ${\bf 1}$, ${\rm var}_i$, ${\rm
low}_i$, ${\rm high}_i$, ${\rm flip}_i)$, with ${\cal V}_i = {\cal X}_i \dotcup
{\cal U}_i = \{x_1, \ldots, x_{20}\} \dotcup \{u_1, \ldots, u_i\}$; thus $n_i = 20$
and $r_i = i$ (note that ${\cal V}_i \subset {\cal V}_j$ for $j > i$);

\fussy

	\item there exists $v_{i} \in V_i, b_{i} \in \B$ s.t.
$\llbracket v_{i}, b_{i}\rrbracket = K_i(\mybold{x}, \mybold{u})$,
being $K_i(\mybold{x}, \mybold{u})$ the COBDD representation of the 
mgo for a {\em buck DC/DC converter with $i$ inputs} (see~\cite{buck-tekrep-2011}
for a description of this system). $K_i$ is an intermediate output of
the \qks \ tool described in~\cite{qks-cav2010}.


\end{itemize}

\begin{table}
  \centering
  \small
  \caption{KSS performaces}\label{expres-table.tex}
  \begin{tabular}{ccccccc}
    \toprule
    $r$ & CPU & MEM & $|K|$ & $|F^{unsh}|$ & $|Sw|$ & \%\\
    \midrule
1 & 2.20e-01 & 4.53e+07 & 12124 & 2545 & 2545 & 0.00e+00\\
2 & 4.20e-01 & 5.29e+07 & 25246 & 5444 & 4536 & 1.67e+01\\
3 & 5.20e-01 & 5.94e+07 & 34741 & 10731 & 8271 & 2.29e+01\\
4 & 6.30e-01 & 6.50e+07 & 43065 & 15165 & 11490 & 2.42e+01\\
    \bottomrule
  \end{tabular}
\end{table}

For each $\rho_i$, we run KSS so as to compute 
\fun{Synthesize}($\rho_i, v_{i}, b_{i}$)  (see Alg.~\ref{synthesize.alg}). In the
following, we will call $\langle w_{1i}, b_{1i}, \ldots, w_{ii}, b_{ii}
\rangle$, with $w_{ji} \in V_i, b_{ji} \in \B$, the 
output of function \fun{SolveFunctionalEq}($\rho_i, v_{i}, b_{i}$) of
Alg.~\ref{actions.alg}. Moreover, we call $f_{1i}, \ldots, f_{ii} : \B^n \to \B$
the $i$ boolean functions s.t. $\llbracket w_{ji}, b_{ji}\rrbracket =
f_{ji}(\mybold{x})$. Note that, by Lemma~\ref{first_phase.lmm}, for all
$\mybold{x} \in {\rm Dom}(K)$, $K_i(\mybold{x}, f_{1i}(\mybold{x}), \ldots,
f_{ii}(\mybold{x})) = 1$. 

All our experiments have been carried out on a
3.0 GHz Intel hyperthreaded Quad Core Linux PC with 8 GB of RAM.

\subsection{KSS Performance}\label{qks-perf.subsec}

In this section we will show the performance (in terms of computation time,
memory, and output size) of the algorithms discussed in Sect.~\ref{algo.tex}.  
Tab.~\ref{expres-table.tex} show our experimental results. The $i$-th row in 
Tab.~\ref{expres-table.tex} corresponds to experiments running KSS so as to
compute \fun{Synthesize}($\rho_i, v_{i}, b_{i}$). Columns in
Tab.~\ref{expres-table.tex} have the following meaning.  Column $r$ shows the
number of action variables, i.e. $|{\cal U}_i|$ (note that $|{\cal X}_i| = 20$
for all $i \in [4]$).  Column {\em CPU} shows the computation time of KSS (in
secs). Column {\em MEM} shows the memory usage for KSS (in bytes). Column $|K|$
shows the number of nodes of the COBDD representation for $K_i(\mybold{x},
\mybold{u})$, i.e. $|V_{v_{i}}|$. Column $|F^{unsh}|$ shows the number of
nodes of the COBDD representations of $f_{1i}, \ldots, f_{ii}$, without
considering nodes sharing among such COBDDs. Note that we do consider nodes
sharing inside each $f_{ji}$ separately. That is, $|F^{unsh}| = \sum_{j = 1}^i
|V_{w_{ji}}|$ is the size of a trivial implementation of $f_{1i}, \ldots,
f_{ii}$ in which each $f_{ji}$ is implemented by a stand-alone C function.
Column $|Sw|$ shows the size of the control software generated by KSS, i.e. the
number of nodes of the COBDD representations $f_{1i}, \ldots, f_{ii}$,
considering also nodes sharing among such COBDDs. That is, $|Sw| = |\cup_{j =
1}^i V_{w_{ji}}|$ is the number of C code blocks generated by
lines~\ref{begin_nodes.algstep}--\ref{end_nodes.algstep} of function
\fun{GenerateCCode} in Alg.~\ref{translate_all.alg}. Finally, Column {\em \%}
shows the gain percentage we obtain by considering node sharing among COBDD
representations for $f_{1i}, \ldots, f_{ii}$, i.e. $(1 -
\frac{|Sw|}{|F^{unsh}|})100$.


From Tab.~\ref{expres-table.tex} we can see that, in less than 1 second and
within 70 MB of RAM we are able to synthesize the control software for the
multi-input buck with $r=4$ action variables, starting from a COBDD
representation of $K$ with about $4 \times 10^4$ nodes. The control software we
synthesize in such a case has about $1.2 \times 10^4$ lines of code, whilest a
control software not taking into account COBDD nodes sharing would have had
about $1.5 \times 10^4$ lines of code. Thus, we obtain a $24\%$ gain towards a
trivial implementation.

\section{Conclusions}\label{conclu.tex}

We presented an algorithm and a tool KSS implementing it which, starting from a
boolean relation $K$ representing the set of implementations meeting the given
system specifications, generates a correct-by-construction C code implementing
$K$. This entails finding boolean functions $F$ s.t. $K(x, F(x)) = 1$ holds, and
then implement such $F$. WCET for the generated control software is at most linear
in $nr$, being $n = |x|$ the number of input arguments for functions in $F$ and
$r$ the number of functions in $F$. Furthermore, we formally proved that
our algorithm is correct.

KSS allows us to synthesize correct-by-construction control software, provided
that $K$ is provably correct w.r.t. initial formal specifications. This is the
case in~\cite{qks-cav2010}, thus this methodology e.g. allows to synthesize
correct-by-construction control software starting from formal specifications for
DTLHSs. We have shown feasibility of our proposed approach by presenting experimental
results on using it to synthesize C controllers for a buck DC-DC converter.

In order to speed-up the resulting WCET, a natural possible future research
direction is to investigate how to parallelize the generated control software,
as well as to improve don't-cares handling in $F$.

\small
\bibliographystyle{plain}
\bibliography{modelchecking}

\end{document}